\documentclass[aps,prb,reprint,twocolumn,showpacs,floatfix,superscriptaddress]{revtex4}
\usepackage{graphicx}
\usepackage{amssymb}
\usepackage{amsmath}
\usepackage{lmodern}
\usepackage{color}

\begin{document}

\title{Energy bands in graphene: Comparison between the tight-binding model
and {\it ab initio} calculations}

\author{E. Kogan}
\email{Eugene.Kogan@biu.ac.il}
\affiliation{Jack and Pearl Resnick Institute, Department of Physics, Bar-Ilan University, Ramat-Gan 52900, Israel}
\author{V. U. Nazarov}
\email{nazarov@gate.sinica.edu.tw}
\affiliation{Research Center for Applied Sciences, Academia Sinica, Taipei 11529, Taiwan}

\author{V. M. Silkin}
\email{vyacheslav.silkin@ehu.es}
\affiliation{Donostia International Physics Center (DIPC), Paseo de Manuel Lardizabal 4, E-20018 San Sebastian/Donostia, Spain,\\
Departamento de F\'{\i}sica de Materiales, Facultad de Ciencias Qu\'{\i}micas, UPV/EHU, Apartado 1072, E-20080 San Sebastian/Donostia, Spain \\
IKERBASQUE, Basque Foundation for Science, 48011 Bilbao, Spain }

\author{M. Kaveh}
\email{Moshe.Kaveh@biu.ac.il}
\affiliation{Jack and Pearl Resnick Institute, Department of Physics, Bar-Ilan University, Ramat-Gan 52900,
Israel}

\date{\today}

\begin{abstract}
We compare the classification of the  electron bands in graphene, obtained  by  group theory algebra in the framework of tight-binding model (TBM), with that  calculated in the density-functional theory (DFT) framework.
Identification in the DFT band-structure of all eight energy bands (four valence and four conduction bands) corresponding to the TBM-derived energy bands is performed and corresponding analysis is presented. The four occupied (three $\sigma$- and one $\pi$-like) and three unoccupied  (two $\sigma$- and one $\pi$-like)  bands given by DFT   closely correspond to those predicted by TBM, both by their symmetry and their dispersion law. However,  the two lowest lying at the $\Gamma$-point unoccupied bands  (one of them of a $\sigma$-like type and the other of a $\pi$-like one), are not of TBM type. According both to their symmetry and to the electron density these bands are plane waves
orthogonal
to the  TBM valence bands; dispersion of these states can be determined unambiguously up to the Brillouin zone borders. On the other hand, the fourth unoccupied band given by the TBM, can be identified among those given by the  DFT band calculations;
it is situated rather high with respect to energy. The interaction of this band with the free-electron states is so strong, that it   exists only in a part of $k$-space.
\end{abstract}


\maketitle

\section{Introduction}

In the course of the study of graphite and a graphite monolayer, called graphene, understanding of the symmetries of the electrons dispersion law in graphene was of crucial importance. Actually, the symmetry classification of the energy bands in graphene (or "two-dimensional graphite") was presented nearly 60 years ago by Lomer in his seminal paper.\cite{lomer} Later the subject was analyzed by Slonczewski and Weiss,\cite{slon} Dresselhaus and Dresselhaus,\cite{dresselhaus} Bassani and Parravicini.\cite{bassani} Some recent approaches to the problem are presented in Refs.  \onlinecite{malard,manes,kogan1,kogan2,elder}.

In the vast majority of papers, studying  the symmetry of the bands, a tight-binding model (TBM) is used. In particular this was done in Ref. \onlinecite{kogan2}, where  the symmetry classification was done by identifying the bands, obtained  in the framework of Density-Functional-Theory (DFT) band structure calculations,\cite{kogan1} with those obtained by applying the group theory algebra to the TBM.
However, the band calculations  give not only the dispersion law, which was used previously, but also the wavefunctions. Moreover, in the DFT band structure of a graphene sheet an additional information about the nearby environment is contained. Thus in such calculations two free-electron-like lowest-energy conduction bands located at energies below the vacuum level with wave functions spatially largely spread into vacuum are observed.\cite{pobaprl83,pobaprl84,wegrprl08} Among these states the lowest-energy band observed experimentally in graphene\cite{papaprl08,arguprl12,nifaprb12} is sharing a common origin with an image-potential state in graphite,\cite{fahiprl83,pobaprl84,pamoprb13} a so-called interlayer band in graphite\cite{holoprb82,fahiprl83,stblprb00} and intercalated graphite,\cite{pobaprl83,cslinp05} image states in nanotubes,\cite{grkrprl02,zawoprl04,scvont10,huzhnl10} and super-atom states in fullerenes.\cite{fezhs08,pezhcr10,dudoprb11} Recently these two states were in
 terpreted as being the DFT analogues of two lowest-energy members of a double-Rydberg series of graphene.\cite{sizhprb09}

In the present work, by comparing the results of the TBM and the DFT approaches to the symmetry
labelling of the energy bands, we identify the eight bands (four valence and four conducting bands) corresponding to all TBM-derived $\sigma$- and $\pi$-like energy bands. The identified conduction bands are all lying, completely or partially, inside the vacuum continuum in the vicinity of the Brillouin zone (BZ) center. However, upon approaching the zone boundaries, these bands experience strong hybridization with the free-electron-like states and dramatically change their spatial localization.

\section{Tight--binding model}

 Partial symmetry analysis of the energy bands in graphene based on group theory algebra in the framework of the TBM was presented in our previous publications \cite{kogan1,kogan2}. This is why in the present work, while briefly mentioning the previously obtained results, we'll concentrate on
 the the symmetry analysis at the point $M$ and the lines $K-M$ and $\Gamma-M$, lacking in our previous publications.

Our TBM space includes four atomic orbitals: $|s,p>$. (Notice that we assume only symmetry of the basis functions with respect to rotations and reflections; the question how these functions are connected with the atomic functions of the isolated carbon atom is irrelevant.)
We look for the solution of the Schr\"{o}edinger equation as a linear combination of the functions
\begin{eqnarray}
\label{tb}
\psi_{\beta;{\bf k}}^j=\sum_{{\bf R}_j}\psi_{\beta}\left({\bf r}-{\bf R}_j\right) e^{i{\bf k\cdot R}_j},
\end{eqnarray}
where $\psi_\beta$ are atomic orbitals, $j=A,B$ labels the sub-lattices, and  ${\bf R}_j$ is the radius vector of an atom in the sublattice $j$.
A  symmetry transformation of the functions $ \psi_{\beta;{\bf k}}^j$ is a direct product of two transformations: the transformation of the sub-lattice functions $\phi^{A,B}_{{\bf k}}$, where
\begin{eqnarray}
\label{2}
\phi_{\bf k}^j=\sum_{{\bf R}_j} e^{i{\bf k\cdot R}_j},
\end{eqnarray}
and the transformation of the orbitals $\psi_{\beta}$. Thus the representations realized by the functions (\ref{tb}) will be the direct product of two representations.

The Hamiltonian of graphene being symmetric with respect to reflection in the graphene plane, the bands built from the $|z>$  orbitals decouple from those built from the $|s,x,y>$  orbitals. The former are odd with respect to reflection, the latter are even. In other words, the former form $\pi$  bands, and the latter form $\sigma$  bands.

The symmetry analysis is natural to start from the most symmetrical  point $\Gamma$.
The group of wave vector ${\bf k}$  at the $\Gamma$ point is $D_{6h}$.
We have to admit that in our previous publications \cite{kogan1,kogan2} we made  mistakes while connecting representations of the group $D_{6h}$ with those of the group $C_{6v}$.
This is why this time we present this transition with maximum details in the Appendix.
There it is shown that at the point $\Gamma$, $|z>$ orbitals realize $A_{2u}+B_{2g}$ representation, $|s>$ orbitals
realize $A_{1g}+B_{1u}$ representation, and $|x,y>$ orbitals realize $E_{1u}+E_{2g}$  representation of the group $D_{6h}$.

The group of wave vector ${\bf k}$  at the $K$ point  is $D_{3h}$. In Ref. \onlinecite{kogan2} it was found that   at this point  the orbitals $|z>$ realize
$E''$
representation, the orbitals $|s>$ realize
$E'$
representation,
and the orbitals $|x,y>$ realize
$A_1'+A_2'+E'$ representation of the group $D_{3h}$.

The group of wave vector ${\bf k}$  at each of the lines constituting triangle $\Gamma-K-M$  is $C_{2v}$.\cite{thomsen}
Representations realized at the $\Gamma$ and $K$ points determine unequivocally representations realized at the lines of the triangle.
\begin{table}
\begin{tabular}{|l|l|rrrr|}
\hline
 $C_{2v}$ &  & $E$ & $C_2$ & $\sigma_v$ &  $\sigma_v'$ \\
& $D_2$ & $E$ & $C_2^z$ & $C_2^y$ &  $C_2^x$ \\
\hline
$A_1;z$ & $A$ & $1$ & 1 & 1 & 1 \\
$B_2;y$ & $B_3;x$ & $1$ & $-1$  & $-1$ & 1    \\
$A_2$ & $B_1;z$ & $1$ & 1 & $-1$ & $-1$ \\
$B_1;x$ & $B_2;y$ & $1$ & $-1$ & 1 & $-1$     \\
\hline
\end{tabular}
\caption{Character table for irreducible representations of $C_{2v}$ and  $D_2$ point groups}
\label{table:d85}
\end{table}

At the line $\Gamma-K$
the symmetry operations for the group $C_{2v}$
correspond respectively to the symmetry operations for the group $D_{3h}$: $C_2- U_2$, $\sigma_v-\sigma$, $\sigma_v'-\sigma_v$; correspond respectively to the symmetry operations for the group $D_{6h}$: $C_2- U_2'$, $\sigma_v-C_2I$, $\sigma_v'-U_2I$. This correspondence allows to obtain compatibility between the one-dimensional representations of the group $D_{6h}$ ($D_{3h}$)
and the representations of the group $C_{2v}$ by inspection.

To obtain the decomposition of the two-dimensional representations of the group $D_{6h}$ ($D_{3h}$) with respect to the representations of the group $C_{2v}$ at the line $\Gamma-K$,
it is convenient to use equation
\begin{eqnarray}
\label{ex}
a_{\alpha}=\frac{1}{g}\sum_G\chi(G)\chi_{\alpha}^*(G),
\end{eqnarray}
which shows how many times a given irreducible representation $\alpha$  is contained in a reducible one.\cite{landau}
In Eq. (\ref{ex}) $g$ is the number of elements in the group, $\chi_{\alpha}(G)$  is the character of an operator  $G$ in the irreducible representation $\alpha$ and $\chi(G)$ is the character of the operator  $G$
in the representation being decomposed. Using Tables  \ref{table:d2} and \ref{table:d85}  we obtain the decomposition of the two-dimensional representations of the group $D_{6h}$ in the form
\begin{eqnarray}
\label{ft}
E_{1u}&=&A_1+B_1\nonumber\\
E_{2g}&=&A_1+B_1,
\end{eqnarray}
and the decomposition of the two-dimensional representations of the group $D_{3h}$ in the form
\begin{eqnarray}
\label{ft2}
E'&=&A_1+B_1\nonumber\\
E''&=&A_2+B_2.
\end{eqnarray}

At the line $\Gamma-M$ the symmetry operations for the group $C_{2v}$
correspond respectively to the symmetry operations for the group  $D_{6h}$: $C_2-U_2$, $\sigma_v-C_2I$,
$\sigma_v'-U_2'I$. This correspondence allows to obtain compatibility between the one-dimensional representations of the group $D_{6h}$
and the representations of the group $C_{2v}$ by inspection.
Using Eq. (\ref{ex})  we again obtain the decomposition of the two-dimensional representations of the group $D_{6h}$ given by Eq. (\ref{ft}).

At the line $K-M$ the symmetry operations for the group $C_{2v}$ correspond respectively to the symmetry operations for the group
 $D_{3h}$: $C_2- U_2$, $\sigma_v-\sigma$,
$\sigma_v'-\sigma_v$. This correspondence allows to obtain compatibility between the one-dimensional representations of the group $D_{3h}$
and the representations of the group $C_{2v}$ by inspection.
Using Eq. (\ref{ex})  we again obtain the decomposition of the two-dimensional representations of the group $D_{3h}$ given by Eq. (\ref{ft2}).

Now consider the $M$ point. The group of wave vector ${\bf k}$  at this point  is $D_{2h}$.
The symmetry analysis of the bands at the point $M$ based on the symmetry of the atomic orbitals in the TBM is presented in the Appendix.
There it is shown that at the point $M$, $|z>$ orbitals realize $B_{1u}+B_{2g}$ representation, $|s>$ orbitals
realize $A_{g}+B_{3u}$, and $|x>$ orbitals realize $A_{g}+B_{3u}$  representation, and $|y>$ orbitals realize $B_{2u}+B_{1g}$  representation of the group $D_{6h}$.

However,  there is another way to find representations realized at the point $M$, based on the compatibility relations. Of course, the two methods are in agreement with each other.
Two groups $C_{2v}$, one at the line $\Gamma_M$, and another at the line $K-M$, being combined,  contain  all  the symmetry operations of the group $D_{2h}$ at the point $M$. Hence representations at the lines $\Gamma-M$ and  $K-M$ being taken together unequivocally determine irreducible representations realized at the point $M$. Such correspondence is presented in Table \ref{table:a3}\cite{thomsen}.
\begin{table}
\begin{tabular}{|c|cccccccc|}
\hline
$M$ & $A_g$ & $B_{1g}$ & $B_{2g}$ & $B_{3g}$ & $A_u$ & $B_{1u}$ & $B_{2u}$ & $B_{3u}$ \\
\hline
$\Gamma-M$ & $A_1$ & $B_{1}$ & $B_{2}$ & $A_{2}$ & $A_2$ & $B_{2}$ & $B_{1}$ & $A_{1}$ \\
$\Gamma-(K)-M$ & $A_1$ & $B_{1}$ & $A_{2}$ & $B_{2}$ & $A_2$ & $B_{2}$ & $A_1$ & $B_{1}$ \\
\hline
\end{tabular}
\caption{Correlation table of the representations of $D_{2h}$, which is the point-group symmetry at $M$, with the representations of $C_{2v}$, which is the point-group symmetry at the lines of the triangle  $\Gamma-K-M$.}
\label{table:a3}
\end{table}

\begin{figure}[h]
\includegraphics[width= \columnwidth, trim=25 0 20 0, clip=true]{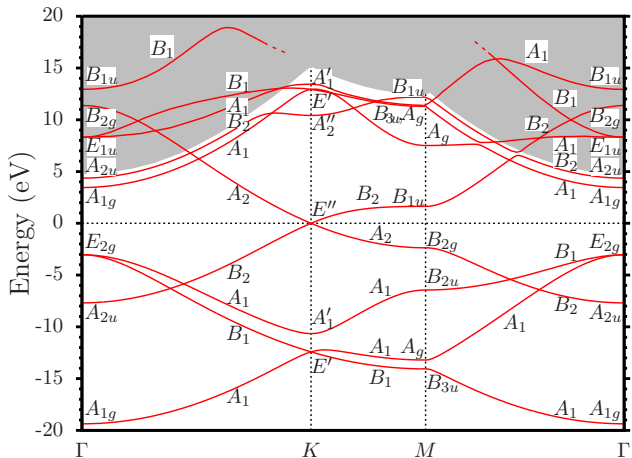}
\caption{\label{fig:bands}  (color online) Graphene band structure evaluated with
use of the FP-LAPW method and the code Elk.\cite{elk} The dashed line shows the Fermi energy. }
\end{figure}

In Fig. \ref{fig:bands} we present the results of the band structure calculations with symmetry labelling of the
valence and the lowest lying conduction bands. Additional mathematical details of the bands  symmetry analysis   will be given in the Appendix.
When looking at the Fig. 1 (and at the Table \ref{table:a3}) it is important to clearly understand the choice of the Cartesian  coordinate systems
(which we did following Ref. \onlinecite{thomsen}). In particular, the principal axis at $M$ is the same as the one at $\Gamma$,
that is the $z$-axis is normal to the plane and the $x$-axis is in the direction to the point $M$.
The Cartesian coordinate system along the $\Gamma-K$ and the $\Gamma-M$ lines
differs from the one at the high-symmetry points \cite{thomsen}. Thus the $z$-axis at the $\Gamma-M$ line is chosen along the $\Gamma-M$ direction.
This explains why, for example, the band which at the $\Gamma-M$ line realizes representation $A_1$ and at the $K-M$ line realizes representation $B_1$,
realizes at the point $M$ representation $B_{3u}$.

Now consider the correspondence between the symmetry of the bands given by the TBM and DFT.
The assumption of the TBM, that is fact that we use the basis consisting of four orbitals per atom with the given symmetry, plus given symmetry of the lattice puts strong restrictions on the symmetry of the electron bands.
The symmetry of all the bonding (valence) bands and the symmetry of the bands, which realize at the $\Gamma$ point representations $E_{2g}$, $B_{2g}$, and $B_{1u}$ obtained from the DFT (see the next Section) corresponds to the predictions of the TBM.

Note that it was recently shown\cite{Nazarov-13} that parts of the
bands inside the vacuum continuum (grey background in Fig. \ref{fig:bands})
turn from true bound-state bands into scattering resonances, by acquiring a finite life-time due to the
coupling of the in-plane and the perpendicular motions. Nevertheless, the current DFT calculation allows us to trace these bands over large portion of the BZ.

\section{DFT band structure}

In this Section we would like to concentrate first on  the two lowest-energy conduction bands,
which are not TBM bands. The lowest conduction  band  has the $A_{1g}$ symmetry at the $\Gamma$ point and $A_1$ along the $\Gamma-K$ line, i.e., it resembles a bonding $\sigma$ state $A_{1g}$ as confirmed by its charge-density distribution in vicinity of the carbon ions presented in Fig. 3 of Ref. \onlinecite{sizhprb09}. As seen in Fig. \ref{fig:bands}, this band maintains an almost free-electron-like character over the entire BZ.
The next conduction band labeled $A_{2u}$ at the $\Gamma$ point, $B_2$ along the $\Gamma-K$ line,  $A_2''$ at the $K$ point, and $B_{1u}$ at the $M$ point, looks like a $\pi$  band. Its charge density distribution around the carbon ions presented at the $\Gamma$ point in Fig. 3 of Ref. \onlinecite{sizhprb09} confirms this assignment.
This band around the $\Gamma$ point also has a free-electron-like dispersion in accordance to the location of the majority of its charge in the vacuum side.\cite{sizhprb09} However, in variance with the lowest-energy conduction band, upon approaching the $K$ point, its dispersion is strongly affected by the interaction with other bands. Thus, charge-density distribution in the $A''_2$ state is strongly attracted to the graphene sheet with the maximum located at $z\approx1.5$ a.u. instead of its location at $z\approx6$ a.u. at the $\Gamma$ point.\cite{pobaprl84,sizhprb09} Moreover, any presence of the $\pi$ component in the vicinity of the carbon ions is washed out in the $A''_2$ state. Upon approaching the $M$ point along the $K-M$ line, the wave function of the states in this band (classified as a $B_{1u}$ state at the $M$ point) re-establishes its large diffusion into the vacuum as seen in Fig. \ref{Charge_M}(c) and the $\pi$-like character around the carbon ions, which is a charac
 teristic of this band at the $\Gamma$ point. The fact that its symmetry is different from that of the three lowest conduction bands can be realized just by looking at the band structure: this band crosses all of them.

\begin{figure}[h!]
\includegraphics[width=0.7 \columnwidth]{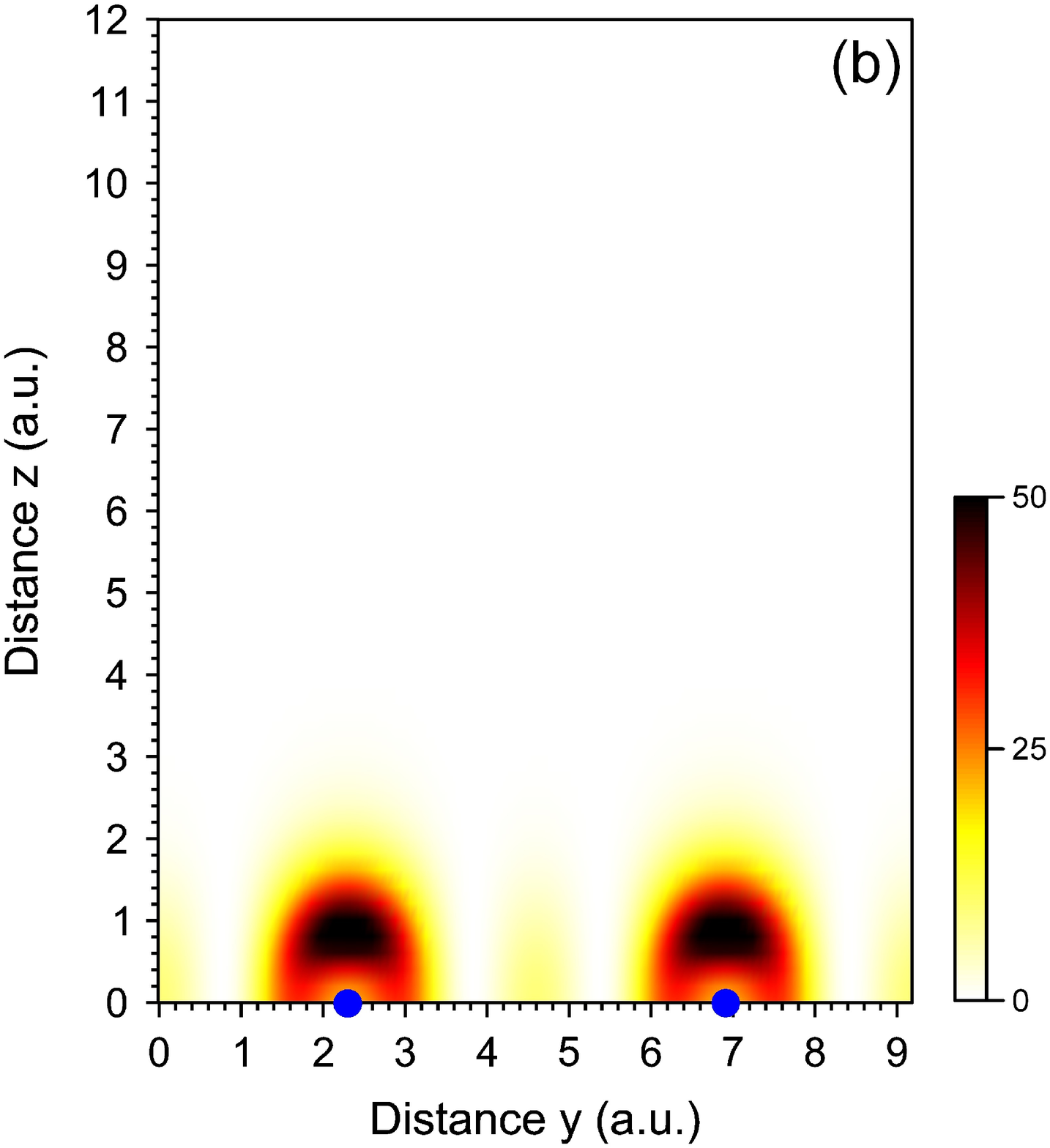}
\includegraphics[width=0.7 \columnwidth]{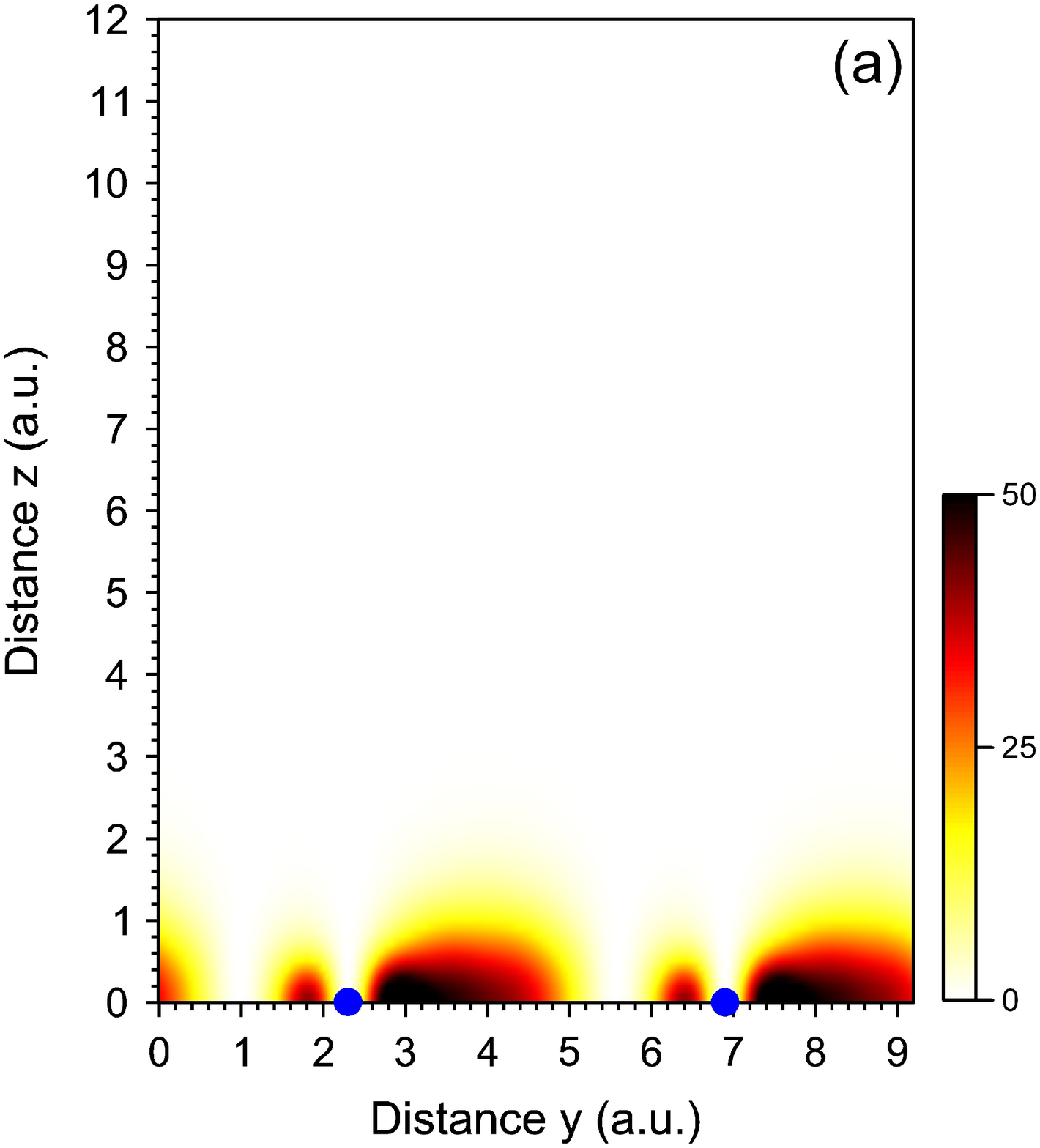}
\caption{\label{Charge_G}  (color online) Charge-density distribution (in arbitrary units) in $x=0$ plane for (a) $E_{2g}$ and (b) $B_{1u}$ states at the $\Gamma$ point. Filled dots show the carbon ion positions.}
\end{figure}

Regarding the other unoccupied energy bands, the most simple situation is with the anti-bonding $\pi$ band which is easily identified and disperses upward from the Fermi level up to energy of +11.4 eV.
On the other hand, the upper-energy band predicted by the TBM has the symmetry  $B_{1u}$ at the $\Gamma$ point and
$B_1$ along the $\Gamma-K$ line. The corresponding band can be identified in Fig. 1 as that having energy of +$13$ eV at $\Gamma$.
The charge-density distribution in this and one of the $E_{2g}$ states is shown in Fig. \ref{Charge_G} where its TBM-like localized character can be easily appreciated.

\begin{figure}[h!]
\includegraphics[width=0.7 \columnwidth]{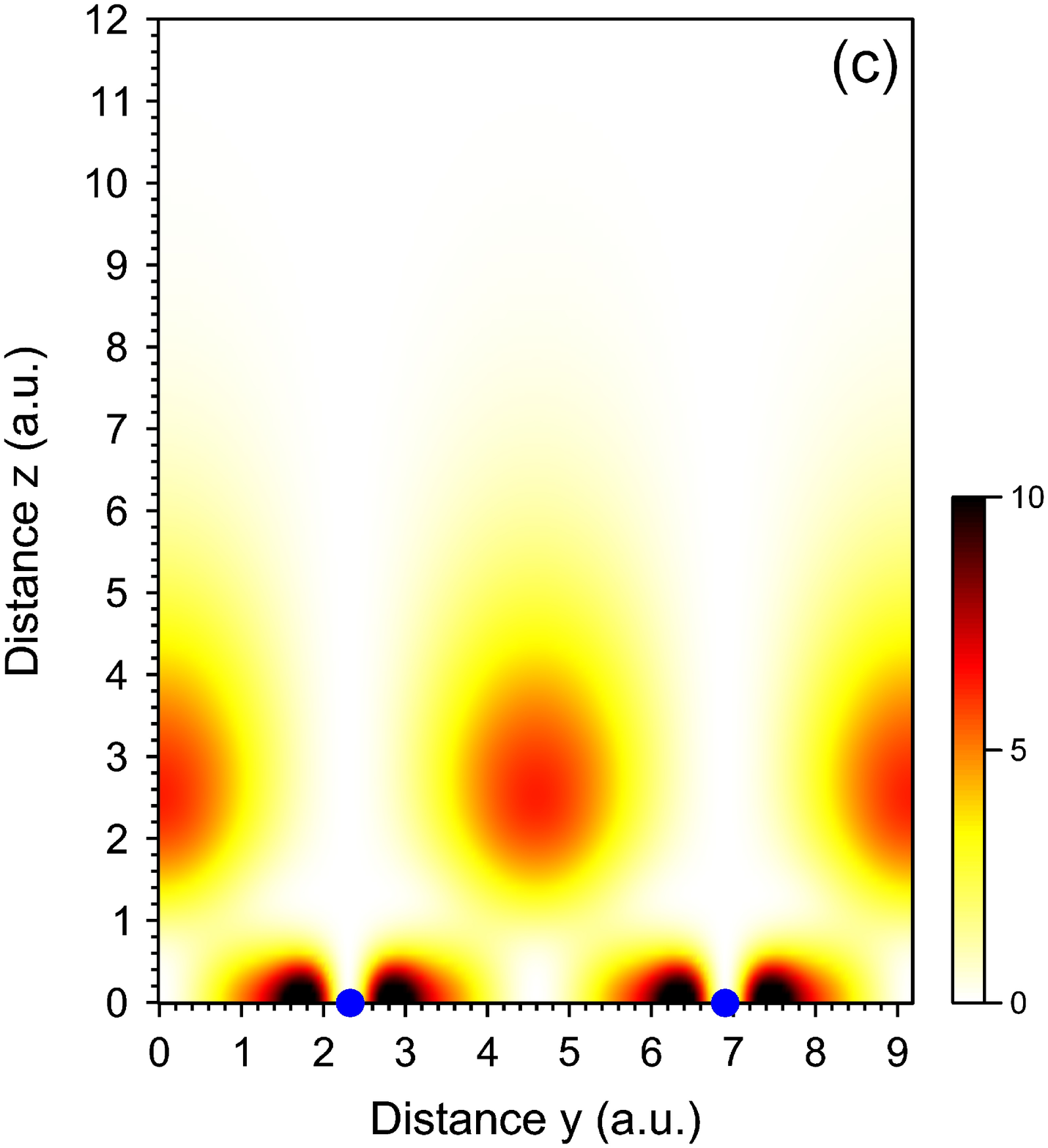}
\includegraphics[width=0.7 \columnwidth]{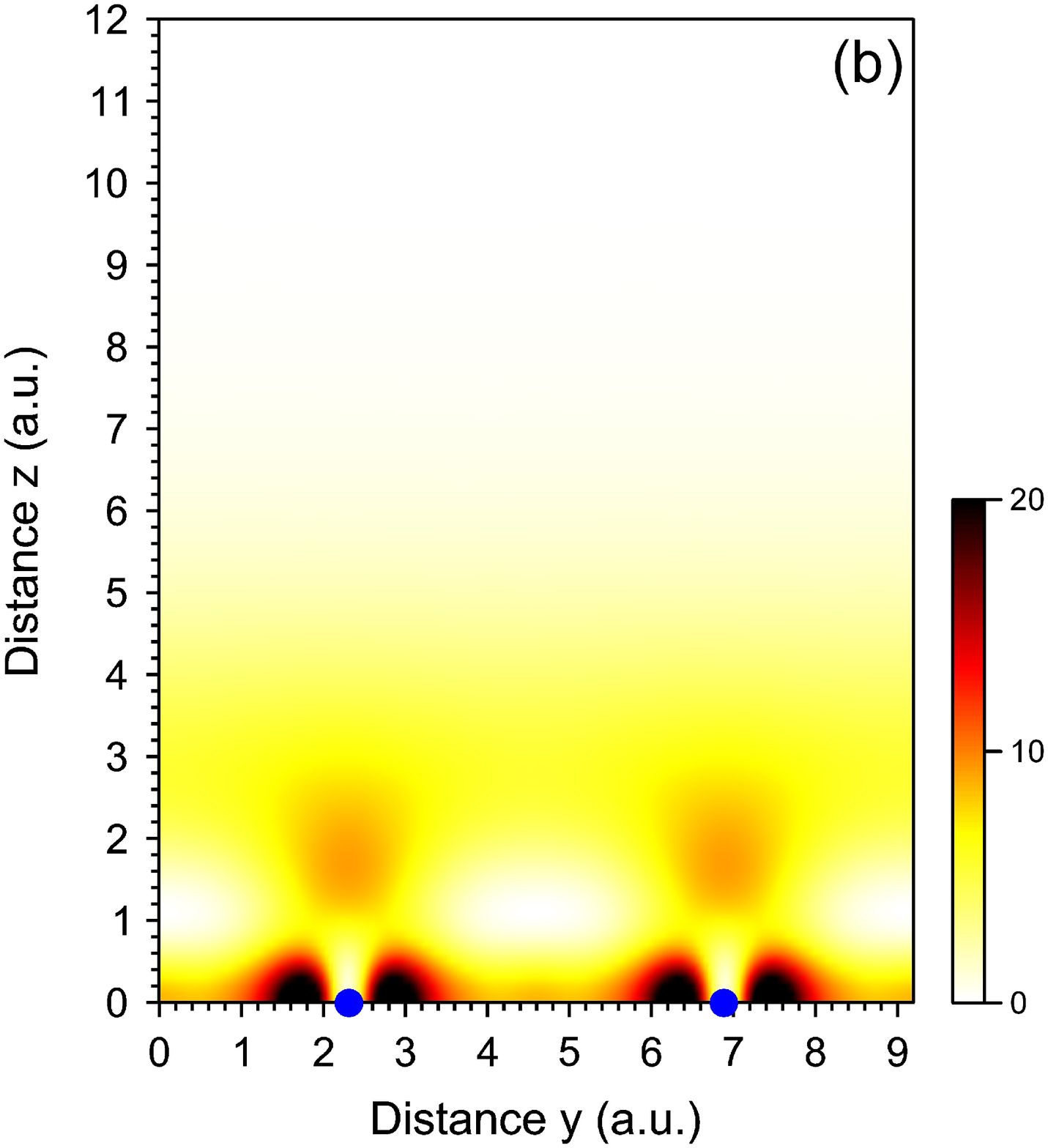}
\includegraphics[width=0.7 \columnwidth]{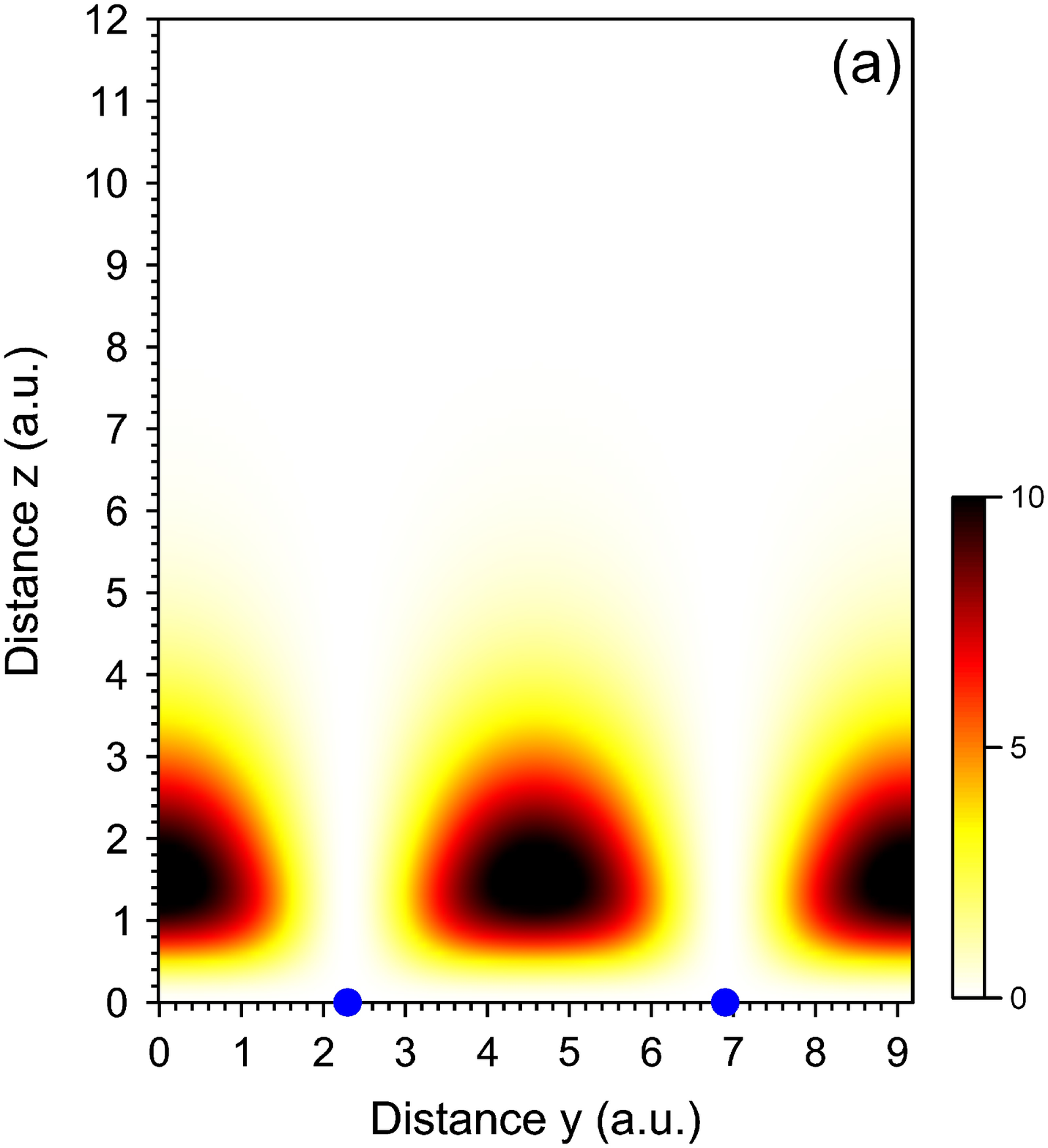}
\label{3}
\caption{\label{Charge_K}  (color online) Charge-density distribution (in Arbitrary units) in $x=0$ plane for (a) $A''_2$, (b) $E'$, and (c) $A_1'$ states at the $K$ point. Filled dots show the carbon ion positions.}
\end{figure}

One of the bands emerging from the double-degenerated $E_{2g}$ state can be traced throughout the whole BZ. Thus in Fig. \ref{fig:bands} it is connected to the $E'$ and $A_g$ states at the $K$ and $M$ points, respectively. As one can note in Fig. \ref{Charge_K}(b) its wave function is distorted from its TBM shape with significant part located on the vacuum side. On the other hand this state at the $M$ point still maintains its atomic-like character as seen in Fig. \ref{Charge_M}(a).

The fate of the second band emanating from the $E_{2g}$ state at finite wave vectors is completely different. Dispersing along the $\Gamma-K$ line it reaches the $K$ point as an $A_1'$ state with its charge-density distribution presented in Fig. \ref{Charge_K}(c). Here one can see that its wave function is even stronger shifted to the vacuum side in comparison with the $E'$ state one. In the $K-M$ direction this band approaches the $M$ point as a $B_{1g}$ state characterized by charge-density distribution reported in Fig. \ref{Charge_M}(b). On the contrary to the situation with the $A_g$ state its wave function has a strong component in the vacuum side. Moving from the $M$ point towards the $\Gamma$ point this band starts quickly dispersing upward. However, reaching the energy around +16 eV this band experiences strong hybridization and slowly disperses downward approaching the $\Gamma$ point. On the contrary, starting from the $\Gamma$ point along t
 he $\Gamma-M$ line this band strongly disperses upward and disappears in the free-electron-like states continuum at energies above $\sim17$ eV.

The upper atomic-like anti-bonding state suffers an even stronger hybridization with the vacuum states continuum. The DFT calculation places this band at the $\Gamma$ point at the energy of $13$ eV (a $B_{1u}$ state). As seen in Fig. \ref{Charge_G}(b), the corresponding
charge-density has a $s$-like symmetry in accordance with the TBM prediction.\cite{bassani} Even being located well-inside the vacuum states continuum this state preserves its atomic-like character in the $\Gamma$-point vicinity. However, moving from the $\Gamma$ point along the $\Gamma-K$ line this band disappears after reaching energy of $\sim19$ eV does not reach the $K$ point. In the $\Gamma-M$ direction this band disperses up to energies about 16 eV from where its dispersion sharply drops down due to hybridization with the free-electron-like states. The dilution of this band within the continuum
finds itself in the perfect agreement with the theory of scattering resonances in 2D crystals,\cite{Nazarov-13}
whereas 2D states above the vacuum level decay due to the coupling between the in-plane and the perpendicular motions.

\begin{figure}[h!]
\includegraphics[width= 0.7 \columnwidth]{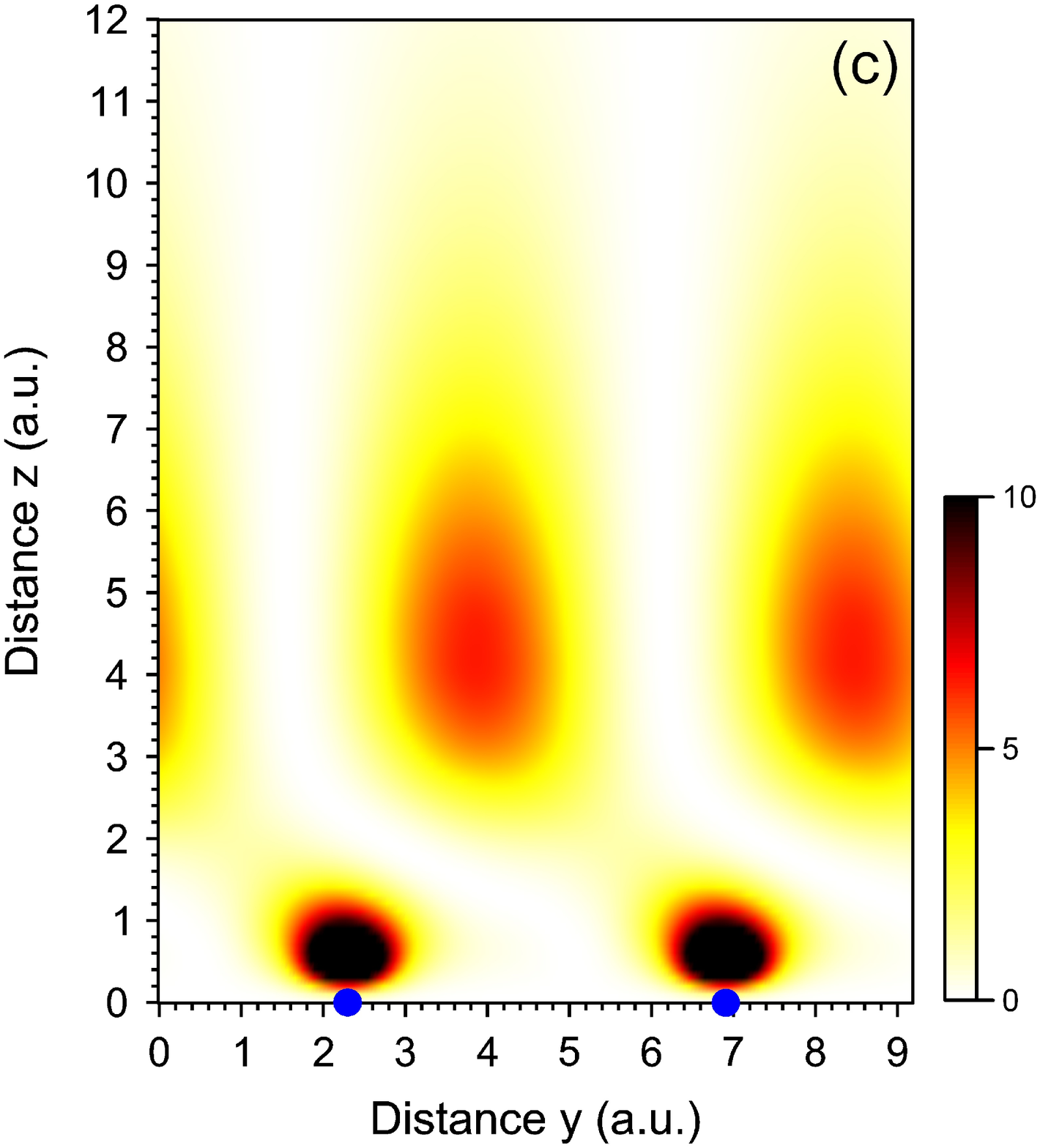}
\includegraphics[width=0.7 \columnwidth]{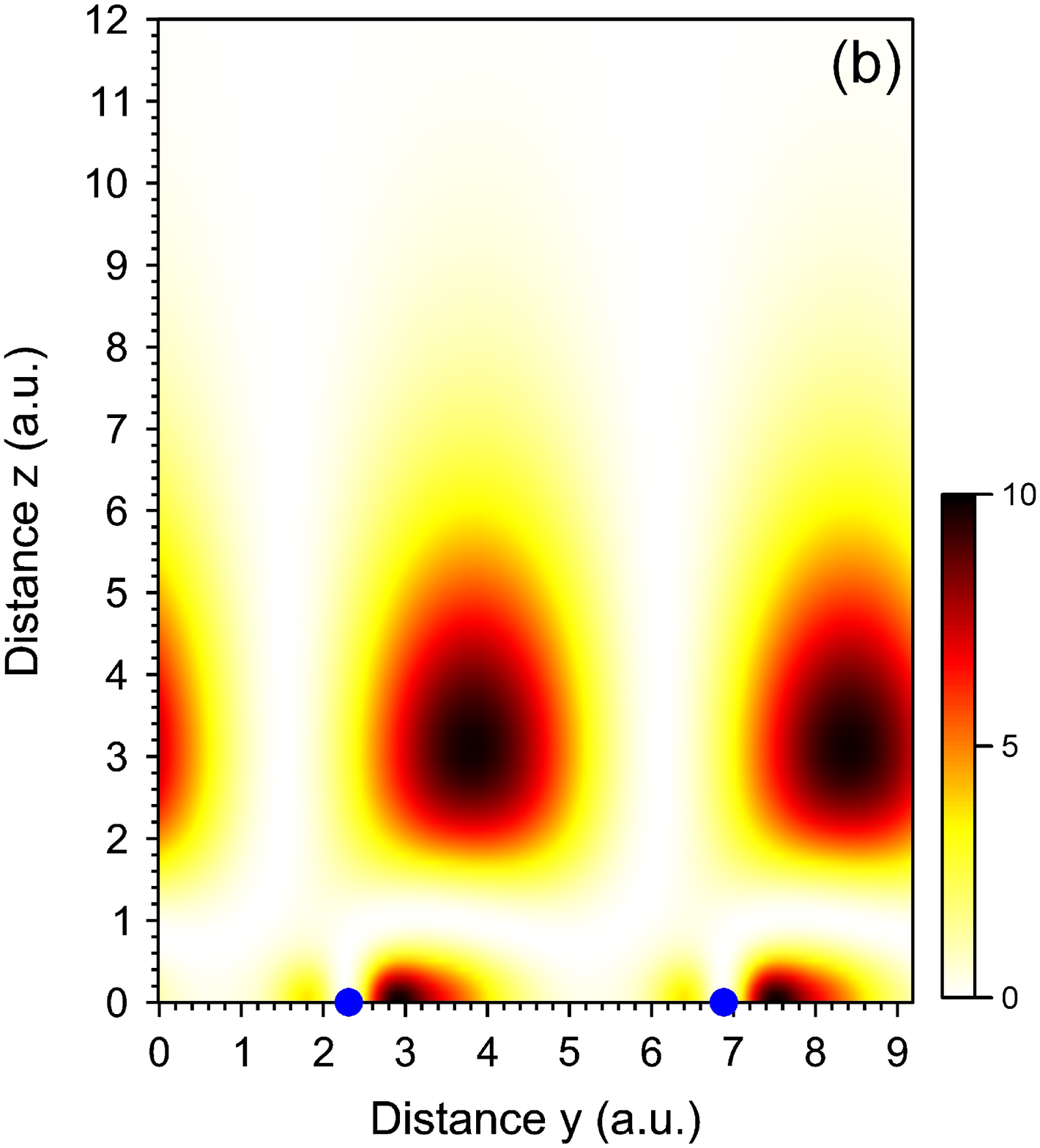}
\includegraphics[width=0.7 \columnwidth]{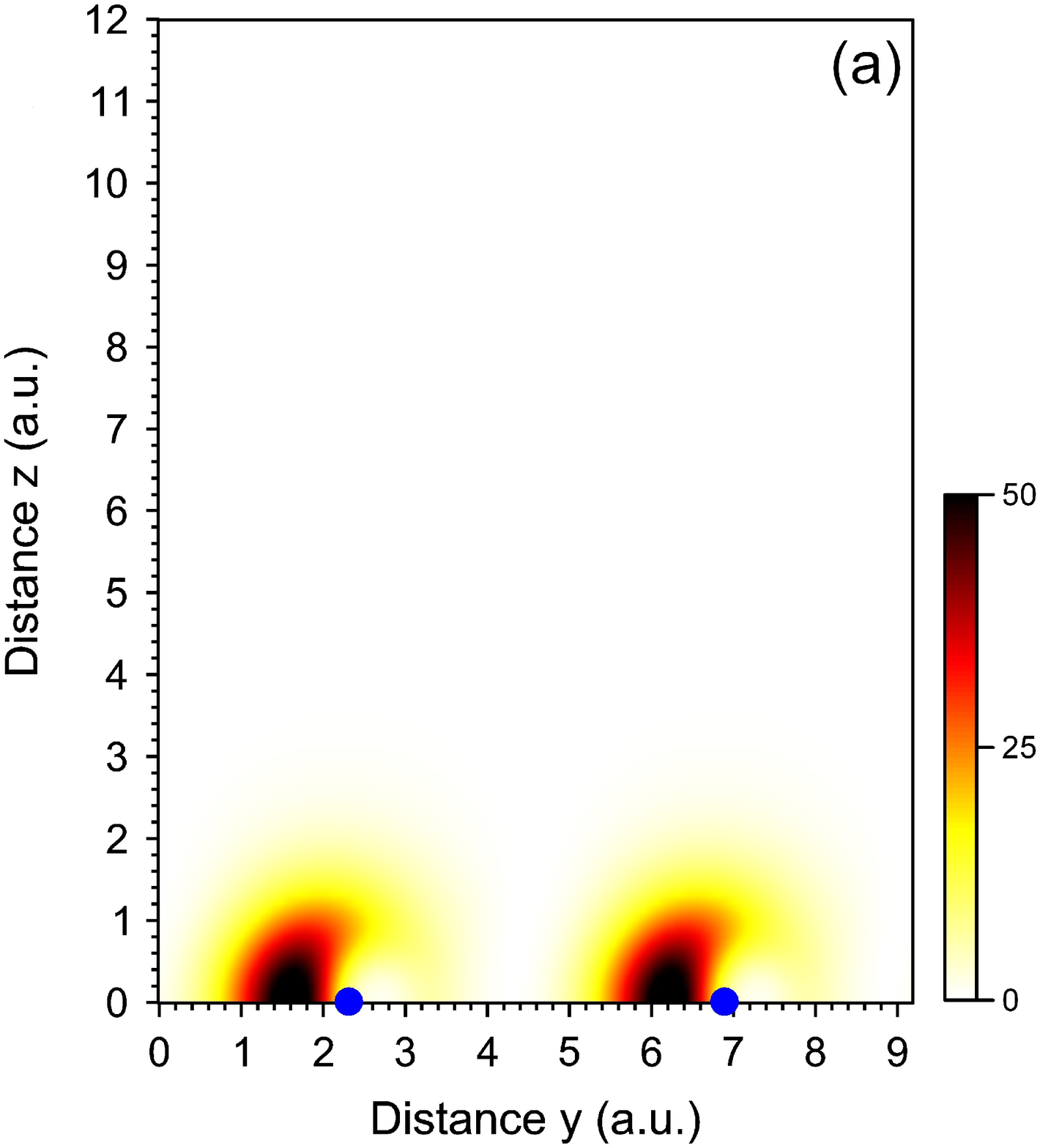}
\label{4}
\caption{\label{Charge_M}  (Color online) Charge-density distribution (in Arbitrary units) in $x=0$ plane for (a) lowest-energy unoccupied $A_g$, (b) $B_{1g}$, and (c) upper-energy $B_{1u}$ states at the $M$ point. Filled dots show the carbon ion positions.}
\end{figure}

\section{Discussions}

First of all we want to explain what does the question in the title of the present work "How good is the tight-binding model?" mean. Group theory algebra
shows that the assumption that   electron wavefunction  can be expanded as a linear combination of four orbitals per atom with the given symmetry, together with given symmetry of the  lattice, unequivocally determines possible representations realized at the symmetry points $\Gamma,K$ and $M$ without any additional assumptions about the Hamiltonian. The question is whether these predictions agree with  the results of the DFT bands calculations.
The answer is that they agree partially. More specifically, all four valence bands and  three out of the five lowest lying conduction bands
obtained by DFT bands calculations correspond to the TBM paradigm.

However,
two lowest (at the $\Gamma$ point) conduction bands given by the DFT band calculation (one of the $\sigma$-type, and another of the $\pi$-type) can not be interpreted in the framework of the TBM. Judging by their symmetry,   these bands can be interpreted as plane waves
(we mean the wave-function dependence upon the $x,y$ coordinates) orthogonal to the bonding bands.
In fact, the lowest energy states built from plane waves will have the maximum  symmetry in the plane $xy$, that is will have the same symmetry
as the lowest bonding bands in the TBM.   Orthogonalization of these plane waves to the bonding bands does not change this
fact.  It is particularly obvious for the non-TBM $\pi^*$ band, because the  plane waves has to be  orthogonal to the band of the maximum symmetry.
The non-TBM $\sigma^*$ band has to be orthogonal to
all three $\sigma$ valence bands, which being taken together, also have the maximum symmetry in the $xy$ plane. And this symmetry is what we see at Fig. 1.

The orthogonal plane wave interpretation of the non-TBM bands is supported also by their
dispersion law and density distribution (see Figs. 3 and 4).

In our previous publications  treating the subject   \cite{kogan1,kogan2},
we started from the dispersion law given by the DFT calculations and essentially equivalent to that presented on Fig. 1. However, no wavefunction analysis was performed in the framework of the DFT calculations, and in the symmetry analysis
we relayed on the group theory exclusively.  Thus the assignment of the irreducible representations was a
delicate process involving the compatibility relations and some guesswork.

Now we have to admit that in our previous publications  \cite{kogan1,kogan2} there were mistakes in labeling the bands.

First, we messed up with the group algebra and  wrongly connected representations of the group $D_{6h}$ with those of the group $C_{6v}$.
This is why this time we present this transition with maximum details.

Second, it is natural to expect that
at the point $\Gamma$ the valence bands are symmetrical with respect to rotation in the plane of graphene by an angle $\pi$ about the center of the line connecting the two
atoms. Such symmetrical combination is said to be bonding \cite{kittel}. Respectively the conduction bands are antisymmetrical with respect to  the rotation (antibonding). If we take into account symmetry of the $|s,x,y>$ orbitals and antisymmetry of the $|z>$ orbitals with respect to reflection in the plane, we come to the conclusion that valence $\sigma$ bands at the point $\Gamma$ should correspond to even representations (index $g$), and the valence $\pi$ band
 should correspond to odd representation (index $u$) \cite{bassani}.
And in our previous publications \cite{kogan1,kogan2} we have ignored this fact while assigning representations to the $|x,y>$ bands at the point $\Gamma$.

Quantitative argument to support the correct assignment was communicated to us by the Anonymous Referee.
In the TBM with the nearest neighbor coupling  at the $\Gamma$ point neglecting overlaps we get
\begin{eqnarray}
E(E_{2g}) - E(E_{1u}) = -3 [H_{pp\sigma} + H_{pp\pi}]
\end{eqnarray}
(see  Ref. \onlinecite{graphene} for notation).
Using the values for the couplings \cite{saito,graphene} $H_{pp\sigma} = 5.1$ eV
and $H_{pp\pi} = -3.1$ eV, we obtain
\begin{eqnarray}
E(E_{2g}) - E(E_{1u}) = -6 eV,
\end{eqnarray}
which means  that  $E_{2g}$ representation characterizes valence band at the point $\Gamma$, and $E_{1u}$ representation characterises conduction band.
And of course this assignment is proved by the  analysis of the wavefunction obtained in the framework of the DFT, which we did.

Analyzing the band structure we have discovered empirically an unexpected topological classification of the bands. There are bands, for which the energy returns to itself when the wave vector changes continuously along the closed curve $\Gamma-K-M-\Gamma$. But there are also bands, where to return to the same value of energy, the wave vector has to traverse the curve two or even three times (see Fig. 1). The more detailed analysis of this classification will be the subject of a separate publication.

Finally, we would like to emphasize that the present paper corrects and extends previous work by some of its
authors.\cite{kogan1,kogan2} The main new contributions are:

-The symmetry classification of energy bands is extended to include the M point and the adjoining lines $K-M$ and $\Gamma-M$ (Fig. 1).

-Some previous assignments of irreducible representations (IRs) are corrected, including those of the two lowest conduction bands (Fig. 1 and Section III).

-The charge-density distributions of some states are presented and discussed (Section III and Figs. 2-4).

\section{Acknowledgements}

Discussions with E. E. Krasovskii, which were very helpful to us, are gratefully acknowledged.

V.U.N. acknowledges  support from National Science Council, Taiwan, Grant No. 100-2112-M-001-025-MY3.

We owe really a lot to the anonymous Referee, who's reports actually corrected several serious mistakes in the initial version of the present paper.

\section{Appendix}

Consider the symmetry analysis at the $\Gamma$ point.
The group of wave vector  is $D_{6h}$. In Ref.  \onlinecite{kogan2}
representations of the group  $D_{6h}$ were obtained on the basis of the identity
\begin{eqnarray}
\label{identity}
D_{6h}= C_{6v}\times C_s.
\end{eqnarray}
It was found  that   the functions   $\psi_{z;{\bf 0}}^j$
realize
\begin{eqnarray}
\label{a}
(A_{1}+B_2)\times A''
\end{eqnarray}
representation, the functions   $\psi_{s;{\bf 0}}^j$ -
\begin{eqnarray}
(A_{1}+B_2)\times A'
\end{eqnarray}
representation,
and the functions   $\psi_{x,y;{\bf 0}}^j$ -
\begin{eqnarray}
\label{c}
(E_{1}+E_{2})\times A'
\end{eqnarray}
representation of the group $D_{6h}$. In Eqs. (\ref{a}) - (\ref{c}) the first multiplier refers to the irreducible representations of the group $C_{6v}$, and the second multiplier refers to the irreducible representations of the group $C_s$ (the character tables are presented in Table \ref{table:d2}).

\begin{table}
\begin{tabular}{|l|l|rr|}
\hline
$C_s$ & & $E$ &  $\sigma$ \\
& $C_i $ & $E$ &  $I$ \\
\hline
$A'$ & $A_g$ & 1 & 1  \\   $A''$ & $A_u$ &1  & $-1$     \\
\hline
\end{tabular}
\begin{tabular}{|l|l|l|rrrrrr|}
\hline
$C_{6v}$ & &  & $E$ & $C_2$ & $2C_3$ & $2C_6$ & $3\sigma_v$ & $3\sigma_v'$ \\
& $D_6$ &  & $E$ & $C_2$ & $2C_3$ & $2C_6$ & $3U_2$ & $3U_2'$ \\
& & $D_{3h}$   & $E$ & $\sigma$ & $2C_3$ & $2S_3$ & $3U_2$ & $3\sigma_v$ \\\hline
$A_{1}$ & $A_{1}$ & $A_1'$ & 1 & 1 & 1 & 1 & 1 & 1 \\
$A_{2}$ & $A_{2}$ & $A_2'$ & 1 & 1 & 1 & 1 & $-1$ & $-1$ \\
$B_{2}$  & $B_{1}$ & $A_1''$ & 1 & $-1$ & 1 & $-1$ & 1 & $-1$ \\
$B_{1}$  & $B_{2}$  &  $A_2''$ &1 & $-1$ & 1 & $-1$ & $-1$ & 1 \\
$E_{2}$  & $E_{2}$  & $E'$ & 2 & 2 & $-1$ & $-1$ & 0 & 0 \\
$E_{1}$  & $E_{1}$ & $E''$  & 2 & $-2$  & $-1$ & 1 & 0 & 0 \\
\hline
\end{tabular}
\caption{Character table for irreducible representations of  $C_s,C_i$  and $C_{6v},D_6,D_{3h}$ point groups}
\label{table:d2}
\end{table}

However, the irreducible representations of the group $D_{6h}$ are traditionally labelled not  on the basis of the identity   (\ref{identity}),
but on the basis of the alternative identity
\begin{eqnarray}
D_{6h}= D_{6}\times C_i.
\end{eqnarray}
Thus each representation of the group  $D_{6}$, say $A_1$, begets two representations: even $A_{1g}$  and odd $A_{1u}$.

To decompose the product of representations (\ref{a}) - (\ref{c}) with respect to the irreducible representations of the group $D_{6h}$ we need
to express the products of the symmetry operations of the groups $C_{6v}$ and $C_s$ through the products of the symmetry operations of the groups $D_{6}$ and $C_i$.
\begin{table}
\begin{tabular}{|c|c|c|c|c|c|c|c|c|c|c|c|}
\hline
$E$ & $C_2$ & $C_3$ & $C_6$ & $U_2$ & $U_2'$ & $I$ & $C_2I$ & $C_3I$ & $C_6I$ & $U_2I$ & $U_2'I$ \\
$E$ & $C_2$ & $C_3$ & $C_6$ & $\sigma_v\sigma$ & $\sigma_v'\sigma$   & $C_2\sigma$ & $\sigma$ & $C_6\sigma$ & $C_3\sigma$ & $\sigma_v'$ & $\sigma_v$ \\
\hline
\end{tabular}
\caption{Correspondence between the products of the symmetry operations of the groups $D_{6}$ and $C_i$ and the products of the symmetry operations of the groups $C_{6v}$ and $C_s$.}
\label{table:d222}
\end{table}
Using elementary algebra we obtain
\begin{eqnarray}
\label{ggg}
\begin{array}{l}
A_{1}\times A'=A_{1g}\\
B_{2}\times A'=B_{1u}\\
A_{1}\times A''=A_{2u}\\
B_{2}\times A''=B_{2g}\\
E_{1}\times A'=E_{1u}\\
E_{2}\times A'=E_{2g}\end{array}.
\end{eqnarray}
All the representations in the r.h.s. of Eq. (\ref{ggg})  are realised at the point $\Gamma$.

Now consider the point $M$. The group of wave vector ${\bf k}$  at the  point  is $D_{2h}$.
Irreducible representations of the point group $D_{2h}$ are obtained on the basis of identity
\begin{eqnarray}
\label{identity2}
D_{2h}= D_{2}\times C_i.
\end{eqnarray}
As it is obvious from Table \ref{table:d85}, the $|z>$ orbitals realize  $B_{1u}$ representation, the $|s>$ orbitals realize $A_g$ representation,   the $|x>$ orbitals realize $B_{3u}$ representation, and the $|y>$ orbitals realize $B_{2u}$ representation of the group $D_{2h}$ (We are considering $M=\left(\frac{2\pi}{3a},0\right)$).
For the basis   $\phi_{\bf M}^j$, we get $\chi(E)=\chi(IC_z)=\chi(C_x)=\chi(IC_y)=2$. The characters corresponding to other transformations are equal to zero. Hence the functions $\phi_{\bf M}^j$ realize $A_g+B_{3u}$ representation of the group $D_{2h}$.
Using elementary algebra we obtain
\begin{eqnarray}
\label{mmm}
\begin{array}{l}
B_{1u}\times A_g=B_{1u}\\
B_{1u}\times B_{3u}=B_{2g}\\
A_{g}\times A_g=A_{g}\\
A_{g}\times B_{3u}=B_{3u}\\
B_{3u}\times A_g=B_{3u}\\
B_{3u}\times B_{3u}=A_{g}\\
B_{2u}\times A_g=B_{2u}\\
B_{2u}\times B_{3u}=B_{1g}\end{array}.
\end{eqnarray}
All the representations in the r.h.s. of Eq. (\ref{mmm}) but one are realised at the point $M$. The missing $B_{1g}$ representation would certainly correspond to the
highest TBA band (see Table \ref{table:a3}), were we able to follow the band to the point $M$.

\end{document}